\documentclass[fleqn,usenatbib]{mnras}

\usepackage{newtxtext,newtxmath}

\usepackage[T1]{fontenc}
\usepackage{hyperref}
\DeclareRobustCommand{\VAN}[3]{#2}
\let\VANthebibliography\thebibliography
\def\thebibliography{\DeclareRobustCommand{\VAN}[3]{##3}\VANthebibliography}

\usepackage{graphicx}	
\usepackage{amsmath}	
\usepackage{romannum}
\usepackage{breqn}

\title[Cygnus A -- Herschel SPIRE]{Cygnus A Obscuring Torus: Ionized, Atomic or Molecular? \thanks{{\it Herschel} is an ESA space observatory with science instruments provided by European-led Principal Investigator consortia and with important participation from NASA.}}

\author[A. Bagul et al.]{
Atharva Bagul,$^{1}$\thanks{E-mail:atharvabagul2000@gmail.com}
Patrick Ogle,$^{2}$
Robert Antonucci,$^{3}$
Philip Maloney,$^{4}$ and
Enrique Lopez Rodriguez$^{5}$
\\
$^{1}$Indian Institute of Science Education and Research Bhopal, 462066 Madhya Pradesh, India\\
$^{2}$Space Telescope Science Institute Baltimore, MD 21218, USA\\
$^{3}$Department of Physics, University of California Santa Barbara, CA 93106, USA \\
$^{4}$Private Astronomer, Boulder, CO 80301, USA \\
$^{5}$Kavli Institute for Particle Astrophysics \& Cosmology (KIPAC), Stanford University, Stanford, CA 94305, USA 
}

\date{Accepted XXX. Received YYY; in original form ZZZ}

\pubyear{2022}

\begin{document}
\label{firstpage}
\pagerange{\pageref{firstpage}--\pageref{lastpage}}
\maketitle

\begin{abstract}
The prototypical powerful FR \Romannum{2} radio galaxy Cygnus A fits extremely well into the quasar/radio galaxy unified model: high polarization with an angle almost perpendicular to the radio jet and polarized flux showing broad permitted lines.
It has been claimed that ionized gas in the torus reveals a very clear torus shape via Bremmstrahlung emission. We rule out the later with an energetic argument, and we constrain the molecular and atomic gas properties with existing observations. 
The atomic absorption against the core has been shown to match the X-ray column only if the spin temperature is an implausible $T_{\rm s} = 1\times 10^6$ K. This points to a molecular medium for the  X-ray column $\log(N_{\rm H} ~[\rm{cm^{-2}}]) \sim 23.5$.  Yet not low-J CO absorption is detected to sensitive limits. The non-detection is surprising given that this powerful radio galaxy hosts a luminous, dust-obscured active nucleus and copious warm molecular hydrogen. These conditions suggest a detectable level of emission. Furthermore, the torus X-ray column density suggests detectable absorption. We explore various possibilities to explain the lack of a signature from warm CO  (200-250K). Specifically, that the radiative excitation by the radio core renders low-J CO absorption below current sensitivities, and that high-J levels are well populated and conducive to producing absorption. We test this hypothesis using archival \textit{Hershel}/SPIRE FTS observations of Cygnus A of high-J CO lines ($14 \geq J \geq 4$ transitions). Still high-J CO lines are not detected. We suggest that ALMA observations near its high frequency limit can be critical to obtain the signature of molecular line of the torus of Cygnus A.
\end{abstract}




\section{Introduction}
\label{sec:intro}
The unification by orientation model of active galactic nuclei (AGN) \citep{antonucci1993unified} posits that the appearance of optical total continuum emission from the central engine (i.e. accretion disk and super massive black hole) depends on the viewing angle with respect to the axis of an obscuring region. This obscuring structure is commonly named as "the dusty torus", but the specific morphology, size, and dynamics are still up for debate. AGN are broadly classify as Type 1 and Type 2. For Type 1 AGN, our line-of-sight (LOS) is close to an unobscured view of the optical continuum emission and broad line region. Those AGN with their central engine emission obscured by an optically thick, molecular, and dusty torus are called Type 2. The obscured broad lines in Type 2 AGN can be seen in polarized flux via scattering by ionized gas or dust across the ionization cones and narrow line regions \citep{antonucci1984optical, antonucci1985spectropolarimetry, barvainis1987hot, miller1991multidirectional}.

While the exact configuration of the torus is still unclear, the torus must provide a large obscuration with respect to its inner radius in order to block roughly $50$ percent of the solid angle near the equatorial plane as seen by the central continuum source \citep[e.g.,][]{elitzur2012}. The torus also must be roughly axisymmetric in order to produce a scattering quasi-conical narrow line region that is photoionized by the central source of ultraviolet (UV) continuum.  Equatorial views within the obscured torus present a high column density of gas, resulting in heavy and sometimes Compton-thick obscuration of the central X-ray source \citep{ghisellini1994contribution}. The torus may be clumpy, with dense clouds of self-shielding dust embedded in a medium of atomic and ionized gas \citep[][for a review]{almeida2017nuclear}. Such an assemblage of dusty molecular clouds might either be supported against collapse by velocity dispersion \citep{krolik1988molecular} or in a molecular outflow ablated from the inner edge of a molecular disk \citep{krolik1986x}. One approach to detect the signature of the obscuring torus is via CO in absorption or emission. CO emission from a compact molecular torus may be undetectable because of the small solid angle and the thermodynamic limit on the brightness temperature. However,
there are not constraints for the strength of CO absorption against a radio core, and thus we can measure CO absorption depending on the geometry, column density of molecular gas, and CO level populations.  An obscuring torus can also be detected by looking for signatures of amorphous/crystalline silicate features \citep{spoon2022}, PAHs \citep{garciabernete2022}, and other sub-mm dense molecular gas tracers like HCO and rotational transitions of vibrationally excited HCN\citep{aaltoHCN2015,falstad2019}.

Cygnus A (3C~405) (z=0.0562) is a classical luminous FR \Romannum{2} galaxy based on its radio morphology and it illustrates the AGN unified model very well. Cygnus A contains a dust-obscured Type 1 AGN revealed by spectro-polarimetric observations that show polarised broad emission lines \citep{ogle1997scattered,1994Natur.371..313A}. The central engine is obscured by a dusty, and potentially clumpy, torus, which is also polarized by an ordered toroidal B-field \citep{2018ApJ...861L..23L}. \citet{carilli1996cygnus} provided a comprehensive review of the multiwavelength properties of Cygnus A. Early interferometric observations revealed Cygnus A as a double radio source \citep{1953Natur.172..996J}. Cygnus A is ten times closer than other radio galaxies of similar radio luminosity ($>10^{45}$ ergs sec$^{-1}$), and contains an AGN with a bolometric luminosity of order $10^{46}$ ergs sec$^{-1}$ comparable to high redshift quasars \citep{runnoe2012updating}.

\cite{barvainis1994search}  put upper limits on the optical depth of CO (1 - 0) absorption against the radio core of Cygnus A. They suggest that the lack of strong CO absorption may be due to the nuclear radio source being several times larger in angular diameter than the individual clouds in the torus, but smaller than the torus itself.

\citet[][hereafter MBR94]{1994ApJ...432..606M} showed the undetected CO absorption can be due to: a) atomic gas, or b) extremely high rotation temperatures and depopulation of the low-J states due to radiative excitation by the bright radio core. Verification of this model would strongly constrain the gas on parsec (pc) scales and imply exotic level populations and absorption spectra. In the region within tens of pc of the compact radio core, the solid-angle averaged brightness temperature would be very high for the low-J transitions. Then, the brightness temperature drops relatively suddenly to values lower than the gas kinetic temperature at J $\sim 5-20$. Qualitatively, high-J levels may be richly populated and thermalized, so that any absorption will not be canceled by stimulated emission.

\cite{ogle2010jet} used Spitzer IRS to detect emission from several $H_2$ pure rotational lines (0-0 S(1), S(3), S(5)-S(7)) in the central 3.7"x3.7" region of Cygnus A.  These $H_2$ emission lines allow to directly trace the molecular gas emission with no $\alpha$CO factor. They use this to estimate warm $H_2$ masses of $1\times10^7 M_\odot$ at 460 K and $5\times10^5 M_\odot$ at 1500 K. They put an upper limit of $< 4\times10^9 M_\odot$ of $H_2$ at T=100 K. Other radio galaxies in their sample have typical cool (50 K) molecular gas mass ratios  of $~0.1$, based on CO detections. Therefore, we might conservatively expect to find $> 1.5\times10^8 M_\odot$ of molecular gas at 50 K  in the central region of Cygnus A  that could potentially be detected in CO emission.  \cite{bellamy2004} also detected hot molecular gas in the dusty structure using NIRSPEC, Keck \Romannum{2} telescope from $H_2$ rotational lines (1-0 S(1), S(2), S(3), S(4)) in the central aperture of PA $105^{\circ}$ data.

Archival \textit{Herschel} Spectral and Photometric Imaging REceiver (SPIRE) \citep{griffin2010herschel}  Fourier-transform spectrometer (FTS) observations provide a unique data set for probing CO emission and absorption in the mid- and high-J lines, and also examine what constraints can be placed on the radiative excitation hypothesis. 

The paper is organised as follows. In Section \ref{sec:observations} we present the observational details of Cygnus A using \textit{Herschel}/SPIRE FTS. The overview of the results obtained from the spectral analysis of the data collected is given in Section \ref{sec:specanal}. In Section \ref{sec:discussion}, \ref{subsec:nondetection} focuses on the failure of the \textit{Herschel}/SPIRE to detect CO emission, \ref{subsec:radiative} discusses whether or not radiative excitation could be important, and \ref{subsec:bremsmodel} comments on why we think the implied recombination luminosity may be incorrect for the bremsstrahlung torus model. Section \ref{subsec:xray_atomic} comments about the failure of X-ray absorption in an atomic torus.

\section{Archival data}
\label{sec:observations}
In the present paper we analyze observations performed with the 3.5-m \textit{Herschel Space Observatory} \citep{2010A&A...518L...1P} using the SPIRE instrument. SPIRE is the \textit{Herschel}'s camera and spectrometer offering three-band imaging photometer operating at $250$, $350$, and $500$ $\mu$m. The FTS uses two overlapping bands to cover 191-671 $\mu$m, the Spectrometer Short Wavelength (SSW) band covers 191-318 $\mu$m and the Spectrometer Long Wavelength (SLW) covers 294-671 $\mu$m. 

We took the publicly available FTS observations from the \textit{Herschel} Archive in sparse sampling mode, with a single pointing of the two coaligned bolometer arrays centered on the nucleus of Cygnus A. This mode is suitable for sources that are smaller than the FTS beam (full width at half maximum, FWHM, $\sim 17\arcsec-42\arcsec$).  For Cygnus~A, $1\arcsec$ corresponds to $1$ kpc, so the instrument beams of the central bolometers cover the bulk of the galaxy at all wavelengths of interest. We retrieved the highly processed products. This special processing, with manual intervention, was applied to isolated, unresolved point sources to optimize background subtraction, utilizing the median background signal from the off-source bolometers. The unapodized data products from these two detectors (SSW and SLW) were used. These two spectra were joined at $1090.64$ and $1090.94$ GHz frequency ($274.88$ $\mu$m and $274.80$ $\mu$m wavelength). The SPIRE integration time was set to make a $5\sigma$ detection at each wavelength in the RJ tail of the warm ($50-100$ K) dust emission. A $3.9$h of exposure was scheduled providing $100$ repetitions, so that the pattern noise limit was not reached. 

The \textit{Herschel}/SPIRE observations were complemented using the following archival observations. Dust continuum Atacama Large (sub-)Millimeter Array (ALMA) observations were taken in four channels centered at 350, 348, 338, and 336 GHz with a 2 GHz bandwidth with a total on-source time of 28.42 minutes with an angular resolution of $0.058$\arcsec, $0.053$\arcsec, and $0.089$\arcsec respectively using ALMA band 7 (ID: 2018.1.01104.S, PI: Perley, Daniel). The phase calibrator was J2007+4029 observed right after the science object. The final image was computed using the standard ALMA pipeline using the Briggs weight and robust = 0.0. The final common beam size is measured to be $0.12\times0.06$ mas$^{2}$ with a position angle of $-3^{\circ}$. The final images reached a sensitivity of 2.7 mJy/beam.\footnote{We do not have information about the sensitivity, except for the comment in the \href{https://almascience.nrao.edu/dataPortal/member.uid___A001_X133d_X29fb.qa2_report.pdf}{QA2} report that the final sensitivity is higher than requested by almost a factor $\sim 10$. As the core of Cygnus A is very bright $(\sim 100 mJy)$, this lost in sensitivity is still enough to obtain reliable fluxes of the continuum emission.} The ALMA data was reduced and analysed via the Common Astronomy Software Applications (CASA) package (CASA Paper cite). CASA is the primary data processing software for the ALMA and NSF's Karl G. Jansky Very Large Array (VLA), and is frequently used also for other radio telescopes. This software can process data from both single-dish and aperture-synthesis telescopes. We downloaded the offline version from their website \footnote{https://casa.nrao.edu} and did the analysis with the help of CASA Documentation \footnote{\href{https://colab.research.google.com/github/casangi/casadocs/blob/3a0ffee/docs/notebooks.introduction.ipynb}{CASA Documentation}} to obtain the results plotted in Figure \ref{fig:sed}.

Dust polarization continuum observations using the Stratopsheric Observatory for Infrared Astronomy (SOFIA) / High-resolution Wideband Camera Plus (HAWC+) at $600$ and $856$ Hz with an angular resolution of $4.85$\arcsec and $7.80$\arcsec, respectively \citep{2018ApJ...861L..23L}. Dust polarization continuum observations using the Gran Telescopio Canarias (GTC) / CanariCam at $1199.2$ and $1873.7$ Hz at an angular resolution of $0.3$\arcsec. For all these observations, Cygnus~A appears as a point-like source.


\begin{figure}
    \centering
    \includegraphics[width=0.5\textwidth]{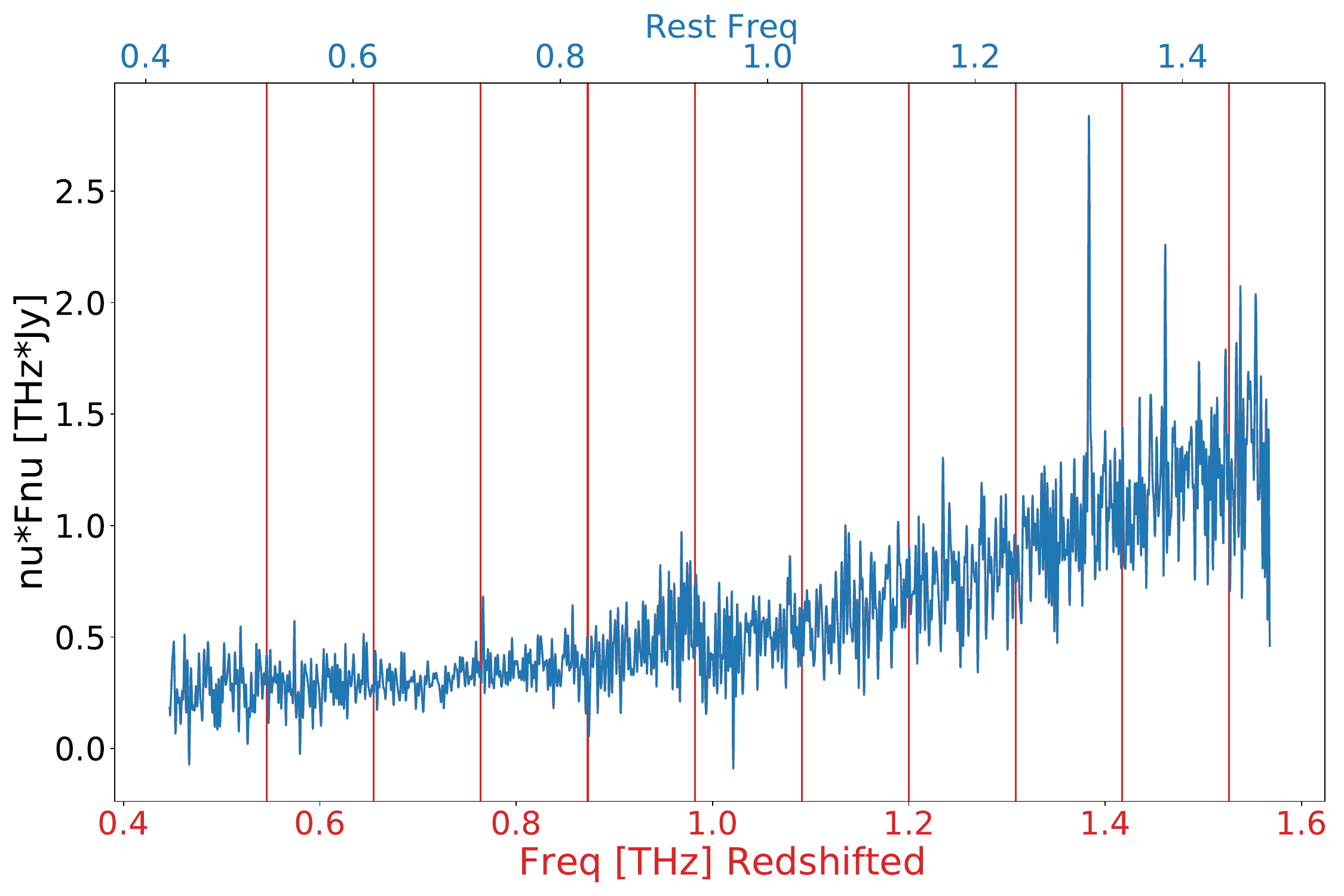}
    \caption{\textit{Herschel}/SPIRE spectrum of CygnusA (3C~405). The $x$-axis is observed frame frequency and the top $x$-axis is the rest frame frequency. The expected frequencies of the C$^{12}$O$^{16}$ rotational ladder from $J=4$ to $J=14$ are shown (red vertical lines).
    at the top x-label needs units.}
    \label{fig:Spectrum}
\end{figure}

\begin{figure}
    \centering
    \includegraphics[width=0.52\textwidth]{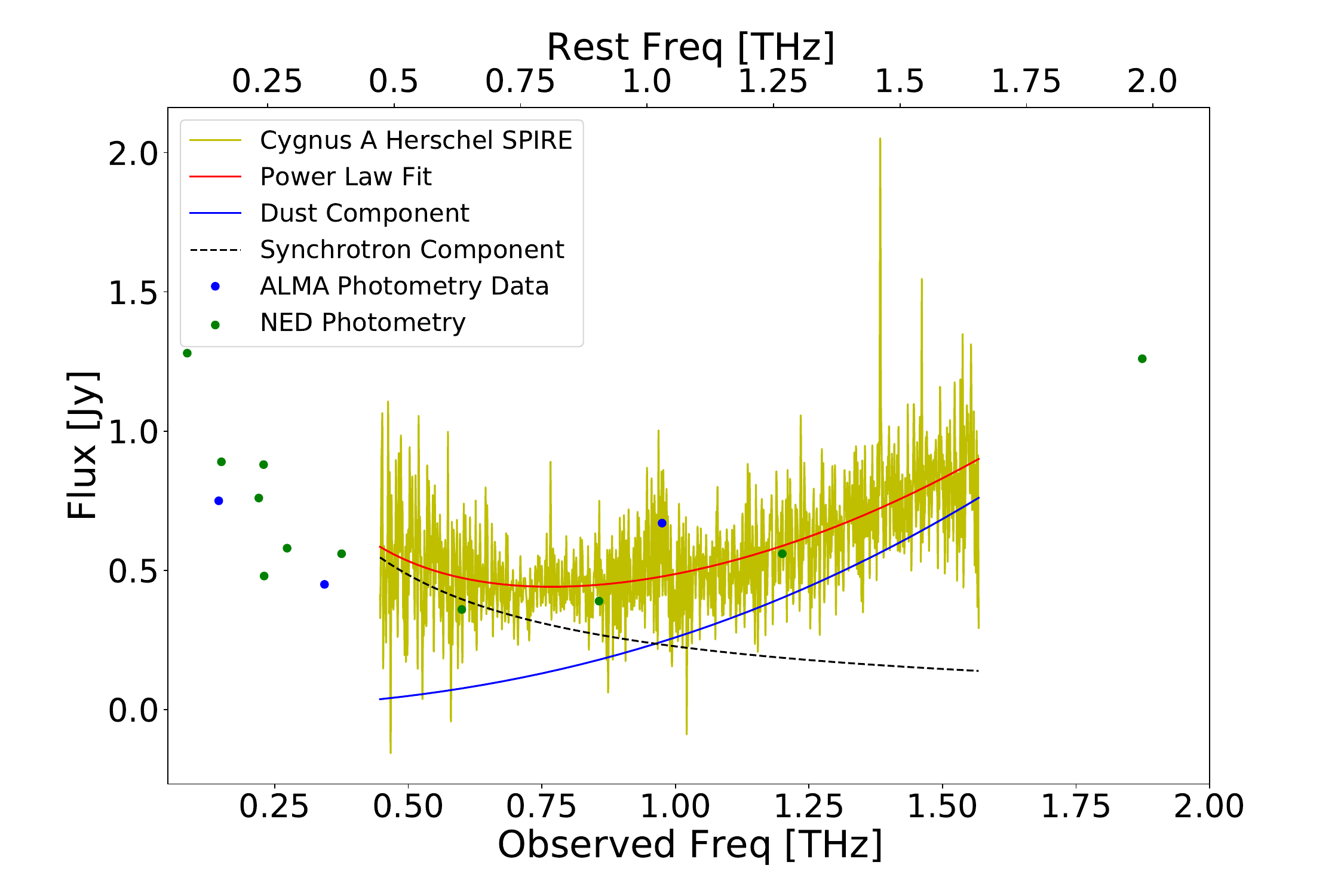}
    \caption{SED of Cygnus A with the best fit model (red solid line) using a dust component (blue solid line) a synchrotron component (black dashed line). The SED fluxes are shown in Table \ref{tab:sed}.}
    \label{fig:sed}
\end{figure}
\section{Spectral Analysis}
\label{sec:specanal}

We show the SPIRE spectrum in Fig. \ref{fig:Spectrum}, with the expected frequencies of the C$^{12}$O$^{16}$ rotational ladder indicated. We found no emission or absorption lines in the $v=0$ state arising from rotational levels from $J = 4$ to as high as $J = 14$.

Figure \ref{fig:sed} shows the $86-1900$ Hz spectral energy distribution (SED) of Cygnus A. The flux densities of these observations are shown in Table \ref{tab:sed} and discussed in Section \ref{sec:observations}. The ALMA fluxes were estimated using the \texttt{CASA} tool \footnote{More information about the tool and its relevant API can be found at the \href{https://casa.nrao.edu/index.shtml}{CASA Website}} \citep{2007ASPC..376..127M}. For these SED values in Table \ref{tab:sed}, we use different apertures. The entire FoV as mentioned in the \ref{tab:sed} is not used as only the nuclear flux is calculated. The flux is concentrated in the unresolved several parsecs scale of the nucleus, so the aperture effects are minimized.
We fit the \textit{Herschel}/SPIRE spectrum using a dust component and a synchrotron component. Both components are assumed to be power-laws in the form 
\begin{equation}
    f(x) = A\left(\frac{x}{x_0}\right)^\alpha
\end{equation} 
where $A$ is the model amplitude, $x_0$ is the reference point and $\alpha$ is the power law index. The two power law functions had $\alpha$ values $2.39$ and $-1.09$ respectively. The final model follows remarkably well the SED of Cygnus A.

Our measurements of the sub-mm continuum and upper limits to the CO line fluxes from the Herschel SPIRE spectrum are given in Table \ref{tab:results}. The expected CO line frequencies are calculated using the rigid rotator approximation and match well the values from \cite{flower2010excitation}; Table 1A).  The density at which collisional de-excitation competes with radiative de-excitation is called the critical density. The critical density for each CO transitions is given by $n_c(H_2)$. $n_c(H_2)$ for CO transition (J to J-1) is approximately $4\times 10^3*J^3$ \citep{kramer04}. Collisional transitions with $\Delta J>1$ are less likely than $\Delta J=1$ and thus the critical density can be approximated as above. The continuum flux values in column 4 (Table \ref{tab:results}) were estimated by taking the mean flux around the line center within a 50GHz-wide window. The standard deviation of the flux within each window is shown in column 6. The instrumental resolution (km/s) for the corresponding transition line center in the last column is calculated from instrumental line width (GHz), that is $1.447$ GHz/$2.355$ as given in the Guide to \textit{Herschel}-SPIRE (version 1.0 Feb 8 2017). We also estimate the $3\sigma$ upper-limit line fluxes as

\begin{equation}
    3\sigma_{\rm lf} = 3\sqrt{2\pi}\sigma_{\rm noise} \theta_{inst,\nu},
\end{equation}
where $3\sigma_{\rm lf}$ is the 3$\sigma$ upper limit to the line flux, $\theta_{inst,\nu}$ is the instrumental resolution for a corresponding line center frequency. These values show the non-detected upper-limit CO absorption or emission in Cygnus A corresponding to the pure rotational transitions.

We estimate the radio core fraction that contributes to the continuum at each CO line frequency (4th column of Table \ref{tab:results} using the two power-law fit shown in \ref{fig:Spectrum}. The lower frequency power-law component comes from the core of the radio jet and that the higher frequency power-law component comes from dust emission. The observed upper-limits to any CO absorption line fluxes and optical depths depend on the covering fractions of these regions by any foreground molecular clouds, which may be different.  For example, if the dust emission is extended over the galaxy, there may be no significant absorption of this component, except by any molecular clouds that lie in dust lanes crossing the nucleus.  Similarly, the absorption of continuum from the radio core depends on its angular extent, which has been resolved at cm wavelengths by the VLA (\cite{carilli2019imaging}; \cite{perley2017discovery}), and the CO cloud covering fraction.  Since these continuum sources are not resolved by \textit{Herschel}, we consider several  cases: 1) both radio and dust emitting sources are compact and covered equally by  CO clouds, 2) only the radio core is compact, 3) only the dust emission is compact, 4) both of the sources are extended and have different covering factors.

We can constrain the geometries of the continuum sources at low frequency using continuum observations retrieved from the ALMA archive (Table \ref{tab:sed}, Section \ref{sec:observations}).  Only the highest frequency (ALMA band 7) overlaps with our \textit{Herschel} observations.  The continuum observed by ALMA is compact with a size of $0.1$\arcsec \footnote{We acknowledge that, given the unresolved nature of the ALMA core of Cygnus A, there is potential for a more precise estimation of its size through the application of advanced deconvolution techniques. These techniques aim to enhance the spatial resolution of the observations, allowing for a more accurate determination of the core's true angular extent. We thus propose that further ALMA observations would be useful.}. The core in ALMA is blended with the jet but it does not probe the dust component and thus we can modify our upper-limits on CO absorption by taking new ALMA observations.

\begin{table}
\resizebox{\columnwidth}{!}{
\begin{tabular}{|c|c|c|c|}
\hline
\begin{tabular}[c]{@{}c@{}}Frequency\\ {[}GHz{]}\end{tabular} & \begin{tabular}[c]{@{}c@{}}Flux Density\\ {[}Jy{]}\end{tabular} & Aperture/FOV & References \\ \hline
86 & 1.28 & - & \citet{agudo2014simultaneous} \\
97.5 & 0.67 (0.22) & 59.732" & ALMA SED (This work) \\
145 & 0.75 (0.25) & 40.163" & ALMA SED (This work) \\
150 & 0.89 & 19" & \citet{monfardini2011dual} \\
220 & 0.76 & 19" & \citet{monfardini2011dual} \\
229 & 0.88 & - & \citet{agudo2014simultaneous} \\
230 & 0.48 & 1.10" $\times$ 0.91" & \citet{wright2004hot} \\
273 & 0.58 & 19" (65mm aperture) & \citet{eales1989photometry} \\
343.5 & 0.45 (0.16) & 16.953" & ALMA SED (This work) \\
375 & 0.56 & 19" & \citet{eales1989photometry} \\
600 & 0.36 (0.11) & 36".29 & \citet{2018ApJ...861L..23L} \\
856.5 & 0.39 (0.09) & 25".67 & \citet{2018ApJ...861L..23L} \\
1199.2 & 0.56 (0.04) & 18".59 & \citet{2018ApJ...861L..23L} \\
1873.7 & 1.26 (0.02) & 10".62 & \citet{2018ApJ...861L..23L} \\ \hline
\end{tabular}
}
\caption{Nuclear fluxes for Cygnus~A (3C~405).}
\label{tab:sed}
\end{table}

\begin{table*}
 \resizebox{\textwidth}{!}{%
\begin{tabular}{ccccccccccc}
\hline
\multicolumn{3}{c}{CO J Transition} & \multicolumn{2}{c}{Continuum} & \multicolumn{5}{c}{Gaussian Fit Profile} & $\theta_{inst, \nu}$  {[}km/s{]} \\ \hline
J & \begin{tabular}[c]{@{}c@{}}Line Center\\ {[}GHz{]}\end{tabular} & \begin{tabular}[c]{@{}c@{}}$n_c(H_2)$\\ {[}$cm^{-3}${]}\end{tabular} & \begin{tabular}[c]{@{}c@{}}Mean Flux\\ (in-band continuum)\\ {[}Jy{]}\end{tabular} & \begin{tabular}[c]{@{}c@{}}Radio Core\\ Fraction\end{tabular} & \begin{tabular}[c]{@{}c@{}}$\sigma$ noise\\ (mean flux)\\ {[}Jy{]}\end{tabular} & \begin{tabular}[c]{@{}c@{}}$3\sigma$ CO\\ Line Flux U.L\\ {[}Jy*km/s{]}\end{tabular} & \begin{tabular}[c]{@{}c@{}}$3\sigma$ CO\\ Line Flux U.L $\times 10^{-18}$\\ {[}W/m\textasciicircum{}2{]}\end{tabular} & \begin{tabular}[c]{@{}c@{}}$3\sigma$ UL\\ to line EW\\ {[}km/s{]}\end{tabular} & \begin{tabular}[c]{@{}c@{}}$L_{CO}$\\ {[}$L_\odot${]}\end{tabular} &  \\ \hline
5-4 & 545.67 & $5\times 10^5$ & 0.523 & 0.798 & 0.183 & 450 & 8.18 & 860 & \textless $1.47\times 10^7$ & 796 \\
6-5 & 654.7467 & $8.6\times 10^5$ & 0.47 & 0.677 & 0.1 & 210 & 4.59 & 447 & \textless{}$8.24\times 10^6$ & 663 \\
7-6 & 763.8121 & $1.4\times 10^6$ & 0.462 & 0.551 & 0.087 & 158 & 4.02 & 342 & \textless{}$7.23\times 10^6$ & 568 \\
8-7 & 872.8426 & $2\times 10^6$ & 0.44 & 0.435 & 0.124 & 196 & 5.72 & 445 & \textless{}$1.02\times 10^7$ & 497 \\
9-8 & 981.8392 & $2.9\times 10^6$ & 0.523 & 0.338 & 0.173 & 236 & 7.74 & 451 & \textless{}$1.39\times 10^7$ & 442 \\
10-9 & 1090.7974 & $4\times 10^6$ & 0.499 & 0.262 & 0.112 & 142 & 5.15 & 285 & \textless{}$9.28\times 10^6$ & 398 \\
11-10 & 1199.713 & $5.3\times 10^6$ & 0.593 & 0.203 & 0.115 & 133 & 5.31 & 224 & \textless{}$9.56\times 10^6$ & 362 \\
12-11 & 1308.581 & $6.9\times 10^6$ & 0.657 & 0.158 & 0.114 & 119 & 5.21 & 181 & \textless{}$9.33\times 10^6$ & 332 \\
13-12 & 1417.399 & $8.8\times 10^6$ & 0.755 & 0.125 & 0.12 & 117 & 5.54 & 155 & \textless{}$9.93\times 10^6$ & 306 \\
14-13 & 1526.161 & $1.1\times 10^7$ & 0.809 & 0.0992 & 0.182 & 164 & 8.35 & 203 & \textless{}$1.50\times 10^7$ & 284 \\ \hline
 &  &  &  &  &  &  &  &  & \begin{tabular}[c]{@{}c@{}}\textless{}$1.08\times 10^8$\\ (Total)\end{tabular} &  \\ \cline{10-10}
\end{tabular}
}
\caption{Measured upper-limit fluxes of the CO$^{12}$O$^{16}$ rotational levels from J=$4$ to J=$14$ from the \textit{Herschel}/Spire spectrum. Columns 1 and 2 are the frequencies in the observed frame corresponding to the J transitions \citep{flower2010excitation}.  Column 3 gives the critical density for each of the transitions. Column 4 shows the measured mean continuum fluxes around the corresponding line center in a $50$GHz-wide window. Radio core fraction calculated for the particular frequency is given by column 5. The features of the Gaussian fit profile are given in columns 6-9. Column 10 gives the CO luminosity in terms of $L_\odot$. Rest of the details are given in Section \ref{sec:specanal}}
\label{tab:results}
\end{table*}

\section{Discussion}
\label{sec:discussion}
\subsection{Non-Detection of CO by \textit{Herschel}/SPIRE}
\label{subsec:nondetection}
As discussed in Section \ref{sec:specanal}, the \textit{Herschel}/SPIRE observations failed to detect any CO lines in emission and absorption.

\subsubsection{Non-Detection of CO in Absorption}
The $3\sigma$ upper limits are expressed as solar luminosities in Table \ref{tab:results}, calculated as

\begin{equation}
L_{\rm CO} = 1.04\times10^{-3} \frac{S_{\rm CO}\nu_{\rm rest}D_{\rm L}^2}{(1+z)}~[L_\odot]
\end{equation}
\noindent
where $S_{\rm CO}$ is measured line flux in units of Jy km s$^{-1}$, $D_{\rm L}=240$ Mpc is the luminosity distance in units of Mpc, and $z$ is the redshift. For all the lines in the \textit{Herschel} data, the upper-limit luminosities are $\lesssim 1\times 10^7 L_\odot$. We estimate a total added upper-limit luminosity of $<10^8 L_\odot$. This upper-limit luminosity is estimated as the sum of the individual line upper-limits. Undoubtedly overestimates the actual upper-limit to the total CO line luminosity. A more careful analysis taking into account that the fluxes in adjacent lines will be correlated (in a manner that depends on the physical conditions in the emitting regions), and the varying noise levels across the \textit{Herschel} spectrum is required. 

Relative to the $J = 8-7$ line luminosity upper limit, the individual line upper limits for Cygnus A range between 0.7 and 1.5. For comparison, in the flux-limited ULIRG sample of \citet{pearson2016}, for the subsample in which all lines from $J = 5-4$ to $J = 13-12$ were detected, these ratios range between 0.2 and 1.6. In comparison, our estimated upper limit to the luminosity over that same ranfe is approximately 10.6 times higher than the upper limit to the J = 8 - 7 line luminosity. However it is important to note that deriving ratios from upper limits can be inconclusive. Given the current data, we made the assumption that all lines within this range possess equal strength, even though this may not be entirely accurate. Co-adding all lines in this manner would yield an upper limit to the flux in any line that would be $\sim \sqrt{10}$ times lower than our present value. Although this approach doesn't account for the variations in noise at different frequencies, it provides a reasonably acceptable estimation.

We then considered an alternative method to calculate the total luminosity by multiplying this improved line flux by $\sim 7$, which is roughly the median ratio observed in the \citet{pearson2016} subsample. The resulting value is about $2.2 \times 10^7$, $\sim 5$ times smaller than the estimate obtained from simply co-adding the observed fluxes. It is essential to acknowledge that there is an inherent uncertainty of at least a factor of 2-3 in this number. Specifically, the estimate could potentially be around 25\% lower or 50\% higher, depending on which lines in Cygnus A are utilized as the reference (in this case, we employed the 8-7 line as the reference for the \citet{pearson2016} subsample). Nevertheless, despite the uncertainties, we believe that this approach represents a reasonably sensible and calculated estimation\footnote{Note that any such attempt to refine the estimate of the luminosity would be model-dependent, as noted above, and as can be seen empirically from inspection of the variability in the CO SLEDs shown in \citet{pearson2016}'s Figure 16.}. 

\subsubsection{Non-Detection of CO in Emission}
Is the failure to detect any emission lines  surprising,  given the expectation based on the presence of a significant amount of hot mass in H2? To answer this question, we need a comparison sample of galaxies with well-characterized CO emission lines in these same mid-$J$ range. The best sample currently available is the set of \textit{Herschel}/SPIRE galaxies analyzed by \citet[][hereafter K14]{2014ApJ...795..174K}. This work established two significant results:

\begin{itemize}
    \item The CO pure rotation lines that fall within the SPIRE passband (effectively, $J = 5\rightarrow 4$ to $J = 14\rightarrow 13$) are produced in a `warm' (more accurately, high-pressure) molecular gas component that contains only a small fraction of the total molecular gas mass but completely dominates the CO line luminosity; and 
    \item The ratio of the total warm CO luminosity to the far-infrared luminosity (the latter defined as the $8-1000$ $\mu$m luminosity) is constant at about $4\times 10^{-4}$, with no discernible dependence on luminosity or galaxy classification (i.e., AGN or starburst).
\end{itemize}

This sample is admittedly limited --- there are only 17 galaxies --- but represents the best currently available. From the obsrved $60~\mu$m flux of Cygnus A, its far-infrared luminosity ($\nu L_\nu$) is $L_{\rm IR} \sim 10^{45}$ erg s$^{-1}$. The K14 result then implies a warm CO luminosity $L_{\rm CO}\sim 10^8 L_\odot$. In other words, if Cygnus A resembles the K14 sample in its ratio of warm CO to far-infrared luminosity (which does not imply that it resembles this rather diverse set of galaxies in any other way), its predicted SPIRE-band CO emission is comparable to the observed $3\sigma$ upper limit. This strongly suggests that the failure to detect any CO  emission line from Cygnus A with \textit{Herschel}/SPIRE is not unexpected. This result does not indicate that there is anything necessarily anomalous about Cygnus A in this respect. Based on the observed far-infrared luminosity, the SPIRE observation is simply not deep enough to have unambiguously detected the emission. 

\citet{carilli2022} detected CO (2-1) emission within -400km/s to +250km/s velocity range, spanning about 4.4kpc in a north-south orientation, closely following the dust lane in HST's (Hubble Space Telescope) I-band imaging (Figure 2 of \citet{carilli2022}). The emission appears clumpy, with two dominant regions approximately 1.5" in size, located northeast and southwest of the nucleus, and faint emission across the galactic center, exhibiting a significant asymmetry between the bright, double-peaked emission in the south and diffuse emission in the north.

\subsection{Radiative Excitation Hypothesis}
\label{subsec:radiative}

To explain the failure of \citet{barvainis1994search} to detect absorption in the $J = 1\rightarrow 0$ line against the radio core of Cygnus A, \citet{1994ApJ...432..606M} suggested that radiative excitation by the nonthermal continuum could increase the excitation temperature of the CO molecules substantially. Thereby, reducing the line optical depths and making the CO column undetectable. Is this suggestion still a viable possibility?, and could radiative excitation play a role in the failure to detect any lines in the SPIRE band?

Thanks to the VLBI study by \citet{2016A&A...588L...9B}, we have a much better understanding of the morphology of the continuum emission at $3$mm than was available in 1994. Most of the flux density, $S_\nu \approx 1$ Jy,  arises within a radius of 0.1 pc, which is much smaller than any size scale likely to be relevant for even a nuclear molecular gas component. We can thus write the mean intensity at the line frequency as

\begin{equation}
    J_\nu \approx 4.6\times 10^{-8} \; {\rm r_{\rm pc}^{-2}\ \; erg\, cm^{-2}\, s^{-1} \, Hz^{-1}\,sr^{-1}}
\end{equation}
where $r_{\rm pc}$ is the distance from the source in pc
The Einstein $A$-coefficient for the $1\rightarrow 0$ line is $A_{10} = 7.165\times 10^{-8}\,{\rm s}^{-1}$. Using the relation between the $A$ and $B$ coefficients, the stimulated emission rate $B_{10}J_\nu$ will exceed the spontaneous emission for any radius $r_{\rm pc} < 1400$ pc, with the ratio scaling as $1/r^2$. Hence even for size scales of 100
-- 200 pc, stimulated emission will completely dominate over spontaneous emission. More importantly, for gas on these size scales the stimulated emission and photon absorption rates will be far larger than the collisional de-excitation rates; e.g., at $r=200$ pc and gas density $n=10^4$ cm$^{-3}$, the stimulated emission rate is an order of magnitude larger than the collisional de-excitation rate for any plausible temperature for molecular gas. Hence it is quite possible that the level populations and thus optical depths in the $1\rightarrow 0$ are significantly affected by radiative excitation by the non-thermal continuum. However, it is possible that radiative excitation could be dominant for the $1\rightarrow 0$ line while having no significant effect on the CO rotation lines within the SPIRE spectrum. This depends crucially on the size scale characterizing the molecular gas.

Implicit in the analysis of MBR94 is the assumption that the solid angle-averaged brightness temperature of the non-thermal continuum is much greater than the gas kinetic temperature at all frequencies of interest. This was in keeping with the then-current paradigm of $r\sim$ pc scales for the obscuring `tori' that block our view of the central engine in Type 2 objects. However, there is abundant evidence for obscuring gas (atomic and molecular) on much larger scales in AGN. In Cygnus A, in particular, the VLBA HI absorption observations of \citet{2010A&A...513A..10S} and the bremsstrahlung torus suggested by \citet{2019ApJ...874L..32C} indicate size scales of $\sim 100-200$ pc for the observed nuclear gas (but see below).  This value is in agreement with those derived from the IR SED modelling with CLUMPY \citep{nenkovaclumpy} torus models \citep[e.g.:][]{privon12}. If this same scale applies to the nuclear molecular gas, then radiative excitation will not be important in the SPIRE spectrum. The reason is because the solid angle-averaged brightness temperatures of the non-thermal continuum will be too low for this process to be important for the warm molecular gas components observable in the SPIRE spectrum.

The slope of the mm/submm-wavelength continuum in Cygnus A is not well determined. However, the slope has only a secondary impact, as we will see. For simplicity, assume the CO rotational line frequencies scale precisely as $J$. Then, we can write the angle-averaged brightness temperature as

\begin{equation}
    T_{\rm b} = \frac{5.53J}{\ln(1 + 4.9\times10^{-7}J^{3+\alpha}r^2_{\rm pc})}
\end{equation}
where the non-thermal continuum flux density goes as $\nu^{-\alpha}$. For the $J = 5\rightarrow 4$ line, the predicted mean brightness temperature is $T_{\rm b} = 4.5\times 10^5$ K and $9.0\times 10^4$ K for $\alpha = 0$ and 1, respectively, for $r_{\rm pc} = 1$. However, if $r_{\rm pc} = 100$, these numbers become 58 K and 20 K. Hence radiative excitation by the non-thermal continuum will not be important on $r \sim 100$ pc or larger size scales.

It is important to note that the scale of the gas producing the large X-ray absorption column has not yet been established. As pointed out by \citet{2010A&A...513A..10S}, although they detect strong, broad ($FWHM = 231\pm 20$ km/s) HI absorption against the counter-jet, which they argue arises in gas at $r \sim 80$ pc, against the core the absorption is very weak. One simple way to resolve this discrepancy, as they also note, is if the X-ray absorbing gas occurs on scales $\ll 100$ pc in a spatially compact torus. In such a torus, the radiative excitation by the continuum could be important for most, if not all, of the CO lines falling within the SPIRE spectrum. This is largely a subject to debate because the mass of molecular gas in the torus would be much too small to have been detected by SPIRE. 

An interesting contrast to Cygnus A is presented by the detection of strong, narrow CO absorption in both the $1\rightarrow 0$ and $2\rightarrow 1$ lines against the bright, compact mm core of the powerful radio galaxy Hydra A at $z = 0.05435$ \citep{2019MNRAS.485..229R,2020MNRAS.496..364R}. The absorption features are very narrow, and appear to arise in a $\sim$ kpc-scale disk of molecular gas. This is precisely the sort of absorption feature that might have been expected in Cygnus A. It is important to note, however, that the integrated {\it emission} in the CO lines in Hydra A is much stronger than the absorption, which raises the question why the observations to date of Cygnus A in the $1\rightarrow 0$ line have not detected any emission, either. 

Does this indicate that there is something anomalous after all about the CO emission in Cygnus A? Perhaps not. The most sensitive current upper limits \citep{2000ApJ...545L.113F, 2005ApJS..159..197E} correspond to a $1\rightarrow 0$ luminosity upper limit of $L'_{\rm CO} < 2.2\times 10^8\;{\rm K\, km\, s^{-1}\, pc^2}$, while the far-IR luminosity $L_{\rm IR} \sim 2-4\times 10^{11} L_\odot$. Plotting this data point in Figure 4 of \citet{2012ApJ...753..135K} and Figure 8 of \citet{2009AJ....138..858C}, who present CO $1\rightarrow 0$ observations of low-redshift Type 2 quasars and ULIRGs, respectively, suggests that Cygnus A may represent one end of a continuum of CO emission for a given IR luminosity, i.e., it represents an extreme, but not an outlier.

\subsection{Bremsstrahlung Torus Model}
\label{subsec:bremsmodel}

The bremsstrahlung torus proposed by \citet{carilli2019imaging} has a radius $R_{\rm t}\sim 264$ pc and a half-thickness $h_{rm t} \sim 143$ pc, yields a total volume $V_{\rm t} = 2\pi^2R_{\rm t} h_{\rm t}^2 \sim 1.1\times 10^8$ pc$^3$. Given the observed emission, they derive a mean electron density of $\bar n_{\rm e} = 490$ cm$^{-3}$. However, \citet{carilli2019imaging} pointed out that this uniformly-filled torus would produce too much Thomson scattering, by at least a factor of 8. Their solution is to make the torus clumpy. The bremsstrahlung emission scales as $n_{\rm e}^2$ and the Thomson optical depth is $\tau_{T}\propto n_{\rm e}$, then the density increases by some factor $\rho$ above the mean value.  Thus, the length scale $L$ decreases through the torus by $\rho^2$, while keeping the emission measure $n_e^2L$ constant, but $\tau_T$ decreases by a factor of $\rho$. 

However, the total mass of ionized gas within the torus raises a serious energetic problem. We first note that the total volume recombination rate within the torus is independent of the clumps. This is most simply seen by noting that the volume recombination rate can be calculated by multiplying the torus area by the column recombination rate, where the latter is fixed by the observations. In other words, we must have $\rho^2\bar n_{\rm e}^2 d = {\rm constant}$, where $d$ is any convenient length scale through the torus. This fixes the effective radius of the torus as $R_{\rm eff} = R_{\rm t}/\rho^2$, and the volume recombination rate is 
\begin{dmath}
    N_{\rm rec} = 2\pi^2 R_{\rm eff} h_{\rm t}^2 \alpha_{\rm B} n_{\rm e}^2 = 2\pi^2 (R_{\rm t}/\rho^2)h_{\rm t}^2\alpha_{\rm B}\rho^2 \bar n_{\rm e}^2 \\
    = \alpha_B\bar n_{\rm e}^2 V_{\rm t}
\end{dmath}
independent of the clumping factor. For $T_{\rm e}=10^4$ K, we then get $N_{\rm rec}\sim 2\times 10^{56}$ s$^{-1}$. Decreasing $T_{\rm e}$ to $8000$ K raises this number by about 20\%. 

Every recombination within the torus releases about one Rydberg of energy (i.e. $2.179\times10^{-11}$ erg s$^{-1}$); if the torus is dusty, all this energy will (given the large torus column density) eventually be absorbed by the dust and re-radiated by the torus. This is a luminosity $L_{\rm rec}\sim 4\times 10^{45}$ erg s$^{-1}$, or $1.1\times 10^{12} L_\odot$.

This is a very large number that exceeds estimates of the total  observed infrared luminosity of Cygnus A by a factor of $3-5$. Note also that this luminosity underestimates the amount of energy that must be radiated by the torus, since it only takes the ionizing luminosity into account, and also ignores the substantial fraction of ionizing photons that will be absorbed directly by the dust rather than the gas. This poses a serious problem for the bremsstrahlung torus interpretation, which can only be alleviated by reducing the size of the torus significantly.

\subsection{The X-Ray obscuration is not produced by an atomic torus}
\label{subsec:xray_atomic}

A possibility raised by MBR94 for the absence of detectable CO absorption is that
the torus is entirely atomic. This is entirely feasible in an X-ray dominated region (XDR) even for such large column densities for the pressure to be below a critical value \citep{1996cyga.book...60M}. However, subsequent observations have ruled this out as a possibility. 

\citet{1995ApJ...449L.131C} detected broad (FWHM $\sim 270$ km s$^{-1}$) HI absorption towards Cygnus A, and argued that the HI absorption could arise in an atomic torus. A later study by \citet{2010A&A...513A..10S} with much improved signal to noise argued that this HI absorption arises in a disk with a radius $r\sim 80$ pc. Notably, strong HI absorption was detected only against the counter-jet; the optical depth against the core is only $\tau = 0.016$, whereas against the counter-jet $\tau > 3$. Such a low optical depth against the core can only be associated with the obscuration measured against the X-ray continuum if the spin temperature is high ($T_{\rm s} ~ 10^6$ K) and the velocity dispersion is very large ($\Delta V \sim 500$ km s$^{-1}$). The former is a plausible value for a nuclear atomic torus within a few hundred pc of the central engine, but the latter requires that the gas is in very close proximity to the central super massive black hole. However, \citet{1995ApJ...449L.131C} pointed out that the radio spectrum is flat down to $1.34$ GHz, implying the emission is still optically thin at this frequency. This places strict limits on the pressure in the torus, which must at minimum equal the absorbed radiation pressure; the fraction of the incident bolometric luminosity that is absorbed can be substantial ($f_{\rm abs}\sim 0.1-0.2$) because of the large column density.

This pressure constraint arises because the free-free absorption depth in an atomic torus is non-negligible due to the significant electron fraction, $x_{\rm e} \sim 3 \times 10^{-2}$, maintained by the high X-ray ionization rate. Scaled to parameters appropriate to Cygnus A (see section 3 of \citet{1996cyga.book...60M}), this constrains the torus pressure to $ P/k \lessapprox 3\times 10^8$ cm$^{-3}$ K. This pressure requires that the torus distance from the nucleus $r\gtrapprox 55$ pc. This is comparable to the scale inferred by \citet{2010A&A...513A..10S} for their HI-absorbing gas, and is completely inconsistent with a gas velocity dispersion large enough to be consistent with the weak HI absorption measured towards the core. Hence we can be confident that the X-ray absorption is not produced in an atomic torus, neither that imaged by \citet{2010A&A...513A..10S} or a putative $r \sim$ pc-scale torus.

\section{Conclusion}
\label{sec:conclusion}

We report the \textit{Herschel}/SPIRE non-detection of CO emission and absorption in Cygnus A, for the $\nu=0$ vibrational state arising from rotational levels $ 14 \geq J \geq 4 $. MBR94 suggested that the optical depth for $J = 1 \rightarrow 0$ transitions is affected by radiative excitation by the non-thermal continuum. However, assuming the $\sim 100-200$ pc scale of obscuring gas found by \citet{2010A&A...513A..10S} and \citet{2019ApJ...874L..32C}, we find that the solid angle-averaged brightness temperature of the non-thermal continuum is too low for radiative excitation to be significant in the \textit{Herschel}/SPIRE spectrum. 

The $3\sigma$ upper limits to the line fluxes of the nuclear Cygnus A \textit{Herschel}/SPIRE spectrum imply a CO luminosity $L_{CO} \lesssim 10^{8} L_\odot $. While this is lower than expected from the median $L_{\rm CO}-L_{\rm FIR}$ relation for galaxies, it falls within the expected range if some of the FIR emission comes from the AGN.  A more accurate estimate requires the knowledge of the size scale of the X-ray absorbing gas, which is unknown. Alternatively, more sensitive ALMA observations will lead to a detection of CO in emission.

Contrary to the suggestion of \citet{2019ApJ...874L..32C}, the sub-mm continuum emission from the torus cannot be attributed to bremsstrahlung.  This conclusion is based on the fact that the resulting recombination luminosity exceeds the estimate of total infrared luminosity for Cygnus A by a factor of 3 to 5. This type of model would only work for sub-mm continuum emission at a much smaller size scale, which may be probed by future ALMA observations of Cygnus A.

\section*{Acknowledgements}
\label{sec:aknowledgments}
RA thanks Dr. David L. Meier and Dr. Masa Imanishi for their helpful suggestions and valuable conversations which helped various parts of this manuscript.

HIFI has been designed and built by a consortium of institutes and university departments from across Europe, Canada and the United States under the leadership of SRON Netherlands Institute for Space Research, Groningen, The Netherlands and with major contributions from Germany, France and the US. Consortium members are: Canada: CSA, U.Waterloo; France: CESR, LAB, LERMA, IRAM; Germany: KOSMA, MPIfR, MPS; Ireland, NUI Maynooth; Italy: ASI, IFSI-INAF, Osservatorio Astrofisico di Arcetri-INAF; Netherlands: SRON, TUD; Poland: CAMK, CBK; Spain: Observatorio Astronómico Nacional (IGN), Centro de Astrobiología (CSIC-INTA). Sweden: Chalmers University of Technology - MC2, RSS \& GARD; Onsala Space Observatory; Swedish National Space Board, Stockholm University - Stockholm Observatory; Switzerland: ETH Zurich, FHNW; USA: Caltech, JPL, NHSC. 

SPIRE has been developed by a consortium of institutes led by Cardiff University (UK) and including Univ. Lethbridge (Canada); NAOC (China); CEA, LAM (France); IFSI, Univ. Padua (Italy); IAC (Spain); Stockholm Observatory (Sweden); Imperial College London, RAL, UCL-MSSL, UKATC, Univ. Sussex (UK); and Caltech, JPL, NHSC, Univ. Colorado (USA). This development has been supported by national funding agencies: CSA (Canada); NAOC (China); CEA, CNES, CNRS (France); ASI (Italy); MCINN (Spain); SNSB (Sweden); STFC, UKSA (UK); and NASA (USA).

Based on observations made with the NASA/DLR Stratospheric Observatory for Infrared Astronomy (SOFIA) under the 05\_0071 Program. SOFIA is jointly operated by the Universities Space Research Association, Inc. (USRA), under NASA contract NNA17BF53C, and the Deutsches SOFIA Institut (DSI) under DLR contract 50 OK 0901 to the University of Stuttgart. 

\section*{Data Availability}

We downloaded the data from \textit{Herschel}/SPIRE data archive \footnote{\url{http://archives.esac.esa.int/hsa/whsa/}}, with the observation id \textbf{1342246994} (PI: Dr. Patrick Ogle) taken in the \textit{SpireSpectroPoint} observing mode. In case any reader has more queries regarding the data availability, they can further contact the author.




\bibliographystyle{mnras}
\bibliography{references} 





\bsp	
\label{lastpage}
\end{document}